\documentclass{nature} 
\usepackage[latin1]{inputenc}
\usepackage[english]{babel}
\usepackage{enumerate}
\usepackage{color}

\usepackage{dcolumn}
\usepackage{bm}
\usepackage{mathptmx}
\usepackage{pifont}
\usepackage[latin1]{inputenc}
\usepackage{cancel}
\usepackage{amssymb}
\usepackage{float}
\usepackage{amsmath}
\usepackage{commath} 
\usepackage{physics}
\usepackage{upgreek}
\usepackage[normalem]{ulem} 
\usepackage{mathrsfs}
\usepackage{xcolor}
\usepackage{graphicx}
\usepackage[normalem]{ulem}
\newcommand\bluesout{\bgroup\markoverwith{\textcolor{blue}{\rule[0.5ex]{2pt}{1.1pt}}}\ULon}
\makeatletter
\let\saved@includegraphics\includegraphics
\AtBeginDocument{\let\includegraphics\saved@includegraphics}
\renewenvironment*{figure}{\@float{figure}}{\end@float}
\makeatother
\title{Unveiling nonlinearities of electromagnetically induced transparency in a THz metamaterial}

\author{Amit Haldar,$^1$ Shriganesh Prabhu,$^2$ and Shovon Pal$^1$}

\begin{document}
\maketitle
\begin{affiliations}
\item School of Physical Sciences, National Institute of Science Education and Research, An OCC of HBNI, Jatni, 752 050 Odisha, India
\item Department of Condensed Matter Physics and Materials Science, Tata Institute of Fundamental Research, 400 005 Mumbai, India
\end{affiliations}

\vspace{-20pt}

\begin{flushleft}
(\today)
\end{flushleft}

\begin{abstract}
Electromagnetically induced transparency (EIT) in terahertz (THz) metamaterials relies on the coherent coupling between a radiative (bright) mode and a subradiant (dark) mode. Understanding the dynamic interplay between the bright and dark modes holds the key to manipulate the mutual interference and hence the transparency. Here, we use nonlinear 2D-THz spectroscopy to scrutinize the dynamics through nonlinearities of the EIT-{\it like} phenomenon in a metamaterial platform that comprises of two coupled resonators. From the temporal profiles of the nonlinear pump-probe and photon-echo signals, we found that the bright mode relaxation time is almost twice the time for the coherent exchange of energy between the two coupled resonators. The multi-peak nature of photon-echo signal and the corresponding temporal signatures further provides a direct visualization of the interference between the dressed states that drives the transparency window in our THz metamaterial. A time-resolved density matrix model accurately describes the observed features, including the cross-peak behavior and the temporal dynamics, establishing the coherent mode coupling as the origin of the transparency window. 
\end{abstract}

\maketitle


\section{Introduction}
Electromagnetically induced transparency (EIT) is a quantum interference phenomenon, originally observed in a laser-driven atomic system~\cite{Harris1997PT}, where the presence of a coherent and intense control (or coupling) light-field dramatically alters the optical properties of a \textit{nonlinear} medium. Under this configuration, the absorption of a weak probe light gets suppressed in the presence of the control field. This leads to a narrow transparency window within the broad absorption spectrum of the probe light. Microscopically, EIT is a coherent optical nonlinear phenomenon that arises due to the destructive interference of two distinct quantum pathways, rendering an otherwise opaque medium transparent within a narrow spectral range around an absorption line~\cite{Harris1997PT,Alzar2002AJP}. This results in a dramatic modification of the medium's dispersion, leading to either fundamentally-relevant observations, such as slow-light effect or technologically-relevant applications as in optical data storage~\cite{Phillips2001PRL,Hau1999Nat}. Historically, EIT finds extensive applications in cold atomic systems~\cite{Hau1999Nat,Shiau2011PRL}, solid-state materials~\cite{Schmidt1996OptCommun,Fan2019PRA}, quantum information processing~\cite{Liu2001Nat,Fleischhauer2002PRA}, nitrogen-vacancy centers in diamonds~\cite{Wei1999PRA,Acosta2013PRL}, and metamaterial devices~\cite{Singh2009PRB,Singh2014APL}. 

While the practical implications of EIT remains crucial for various quantum technological applications, its requirement for either cryogenics or vapor cells and bulky laser-vacuum setups often makes it less favorable for compact, on-chip integration. Such evident challenges have prompted the development of EIT analogues using alternative platforms. Here, classical systems such as coupled resonators~\cite{Xu2006PRL,Yang2009PRL}, electrical circuits~\cite{Alzar2002AJP}, and plasmonic structures~\cite{Singh2009PRB,Gu2012NatCom} that exhibit EIT-\textit{like} response have drawn tremendous attention. In particular, metamaterial-based EIT analogues have gained significant prominence due to their tunability, scalability, and potential to operate across a broad frequency spectrum. The unique ability of metamaterial structures to manipulate electromagnetic waves opens new possibilities for exploring EIT-\textit{like} phenomena in technologically-important frequency spectrum such as the terahertz (THz) domain. In the THz metamaterial system, the EIT-\textit{like} spectral behavior arises primarily due to the near-field coupling between the bright-dark modes~\cite{Singh2009PRB,Singh2014APL,Gu2012NatCom,Manjappa2017OL,Limonov2017NP} or the bright-bright modes~\cite{Zhang2017OC,Yahiaoui2018PRB}. The bright mode, characterized by a low quality factor due to its superradiant nature, contrasts the dark mode that has a high quality factor due to its subradiant nature. The destructive interference between these modes results in the characteristic transparency window within a broad absorption spectral bandwidth, closely resembling the quantum EIT phenomenon.

Over the past decade, extensive research has shown that it is not just possible to achieve a single transparency window~\cite{Singh2009PRB,Singh2014APL, Manjappa2017OL,Zhang2017OC,Yahiaoui2018PRB}, but with careful design architectures of meta-structures, one can achieve either a dual transparency~\cite{Vaswani2024OLT, Haldar2025AQT, Hu2024OC, Devi2018JAP, Liu2012APL} or even multiple transparency windows~\cite{Wang2023NC}. Such design architectures range from asymmetric split-ring resonators~\cite{Burrow2017OptExp} and coupled cut-wire bar systems to toroidal~\cite{Zeng2025NC,Bhattacharya2021SciRep,Shu2024OQE} and polarization-sensitive~\cite{Zeng2025NC,Yang2021AP,Cheng2020AA,Gao2022OptLT,Burrow2017OptExp} designs -- allowing precise control over spectral tunability. Polarization-dependent EIT studies have further revealed coupling-like responses and tunable transparency windows~\cite{Wang2024JPAP}, enabling functionalities such as polarization-selective switches~\cite{Lu2025RP} and slow-light manipulation~\cite{Gao2022OptLT,Burrow2017OptExp}. By integrating photosensitive semiconductors such as silicon~\cite{Hu2023ACSPh,Gu2012NatCom,Hu2020LPR}, gallium arsenide~\cite{Padilla2006LPRL,Yan2023OME,Deng2021AMI}, or germanium~\cite{Lim2018AdvMa,He2021PhRe} with metallic or dielectric meta-atoms, active metamaterial platforms have been developed that enable optical excitation of carriers to tune the resonance frequencies or linewidths and/or transmission amplitudes on picosecond timescales. These optically-reconfigurable THz-EIT metamaterials have been shown to demonstrate remarkable functionalities such as ultrafast switching~\cite{Gu2012NatCom,Hu2020LPR,Padilla2006LPRL,Lim2018AdvMa}, transient transparency control~\cite{Hu2020LPR,Yan2023OME}, and modulation-depth enhancement~\cite{Huang2022Opt,Zhoua2020Naph,Hu2020LPR, Yan2023OME,Lim2018AdvMa}, under all-optical excitation. For seamless optimization of such functionalities, it is important to gain dynamic all-optical control and ultrafast modulation of the EIT-like responses under non-equilibrium settings. In this context, optical pump-THz probe (OPTP) spectroscopy has been extensively used to address the mechanisms underpinning the optical driving of the EIT responses~\cite{Devi2022NJP,Xu2023Sen,Hu2023ACSPh}. 

In all experimental implementations involving THz-EIT metamaterials that have been reported thus far, the independent or stand-alone contributions of the metamaterial in an optically-excited EIT system are yet to be addressed. It is to be noted that the optical pump predominantly excites charge carriers within the substrate~\cite{Hu2023ACSPh,Lim2018AdvMa} or the integrated material~\cite{Deng2021AMI,Hu2021ACSPh}, rather than the THz-metamaterial resonators themselves. Consequently, the measured transient THz response represents a convolution of the substrate dynamics that is facilitated by the metamaterial resonators. The OPTP spectroscopy is innately more sensitive to linear or weakly nonlinear responses and hence cannot capture the complex coherent interactions that govern the coupled bright and dark modes responsible for the EIT-\textit{like} behavior. The observed dynamics, therefore, reflect collective effects arising from the entire hybrid system rather than from the metamaterial architecture alone. To the best of our knowledge, no experimental study has yet achieved a truly metamaterial-only THz-EIT response, free from substrate influence, in the presence of a resonant excitation pulse, which in this case is the THz pulse itself. Advanced nonlinear spectroscopic approaches, such as two-dimensional (2D) coherent THz spectroscopy has emerged as a powerful alternative that can be used to address the non-equilibrium THz-EIT dynamics~\cite{Dutta2025JPCM,Liu2025nQM, Kuehn2009JCP}. The 2D-THz spectroscopy would directly capture the EIT dynamics in the non-equilibrium regime, enabling access to coherent coupling, energy exchange, and dephasing processes between the interacting bright and dark modes in the EIT metamaterial without exciting the carriers from the substrate underneath. In particular, 2D THz spectroscopy provides the multi-dimensional correlation map that disentangles homogeneous and inhomogeneous broadening effects~\cite{Salvador2024PRB,Liu2024NatPhys}, offering direct insight into the mode coupling strength~\cite{Kuhen2011JPCB,Kuhen2011PRL}, coherence lifetimes~\cite{Raab2019OptExp,Markmann2021NaPh,Dutta2025AFM,Wan2019PRL}, and nonlinear interactions channels~\cite{Yang2023NRM,Dutta2025JPCM} that technically remain inaccessible in conventional OPTP measurements.

In this article, we drive a THz-EIT metamaterial system in the nonlinear regime of light-matter interaction using the 2D-THz spectroscopy and explore the EIT dynamics at picosecond timescales. We use a time-resolved density matrix model, where the elements of the full density matrix are evolved according to the Liouville-von-Neumann equation in the Lindblad form to model the nonlinear response of the EIT-\textit{like} phenomenon. We show that the THz excitation pulse perturbs the coupled-resonator system, generating transient nonlinear signals that categorically trace how the induced currents evolve and decay over time. The temporal decay of these nonlinear components provides the  relaxation time of excited state in the EIT-\textit{like} system, reflecting how long the resonant current persists before dissipating. From the four-wave-mixing echo signals, we have extracted the coherence timescales between the bright and the dark modes. We also show that the system Hamiltonian very precisely reproduces the cross-peak behavior of the echo signals along with the coherence decay. Our experimental observations, substantiated by the theoretical modeling affirms the mode-resolved picture of the sole EIT behavior in a THz metamaterial -- the transparency window arising purely from coherent interference between resonant currents within the metamaterial. These insights in to the light-matter interactions underpinning the THz-EIT metamaterial paves the way for a better control on the coherence dynamics, towards improved technologically-relevant applications.

\section{Results and discussion}
\begin{figure}
    \centering
    \includegraphics[width=0.8\linewidth]{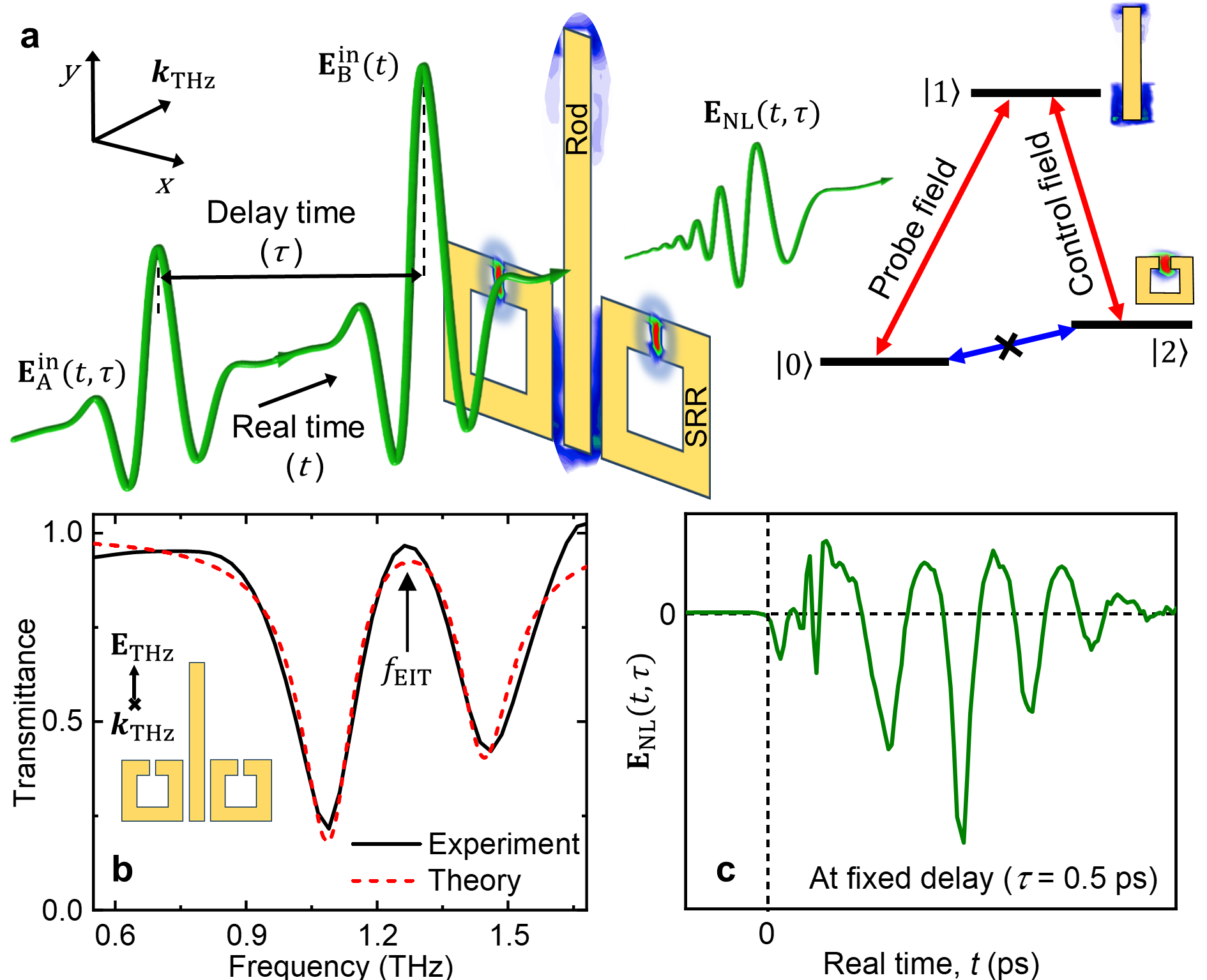}
    \caption{{\bf (a)} {\bf Left}: Schematic representation of the experimental geometry for two THz field interactions with the EIT-{\it like} metamaterial. Here, $t$ is the real time and $\tau$ is the delay time between the two THz pulses ${\bf E}^{\rm in}_{\rm A}(t,\tau)$ and ${\bf E}^{\rm in}_{\rm B}(t)$. ${\bf E}_{\rm NL} (t,\tau)$ is the emitted nonlinear signal. The EIT-{\it like} metamaterial unit cell consists of a central rod resonator with a pair of split-ring resonator (SRR) positioned symmetrically on either side. {\bf Right}: A schematic of the 3-level $\Lambda$-type system used to describe the EIT phenomenon. {\bf (b)} The experimentally measured (black-solid curve) and theoretically modeled (red-dashed curve) transmittance spectra of the EIT-{\it like} THz metamaterial, showing beautiful agreement, with the EIT peak appearing at $f_{\rm EIT} = 1.27$\,THz. The inset shows the incident THz electric-field polarization relative to the metamaterial structure. {\bf (c)} A typical temporal scan of the measured ${\bf E}_{\rm NL} (t,\tau)$ showing coherent oscillations at a specific delay time of 0.5\,ps.}
    \label{fig1}
\end{figure}

We choose a metamaterial unit cell that comprises of a pair of split-ring resonators (SRRs) symmetrically positioned on both sides of the dipole-like central rod, shown schematically in Figure~\ref{fig1}a. To realize the bright-dark mode EIT-{\it like} response, the rod and SRRs operates within the near-field coupling regime~\cite{Singh2014APL,Vaswani2024OLT} that has been established recently for this metamaterial design architecture~\cite{Haldar2025AQT}. In this configuration, the rod exhibits a dipole resonance $f_{\rm rod}$ under $y$-polarized THz pulse excitation, thereby acting as the bright mode for our THz-EIT system. Owing to structural symmetry, the SRRs cannot be directly excited by the same $y$-polarized THz field. In contrast, the SRRs display an LC-type resonance mode $f_{\rm SRR}$ under $x$-polarized THz excitation, serving as the dark mode for our THz-EIT system under the $y$-polarized THz pulse excitation. Evidently, even in the absence of an external $x$-polarized THz field, the SRR gets indirectly driven by the rod's THz near-field distribution~\cite{Haldar2025AQT}. This leads to a destructive interference between the dipole fields of the rod and split-gap fields of the SRR. The interference thereby suppresses the rod's resonance and opens a transparency window, resulting in an EIT-{\it like} response. Figure~\ref{fig1}b shows the experimentally-measured transmittance spectra (black-solid curve) of the THz-EIT meta-structure, which clearly shows a transparency window centered at $f_{\rm EIT}$ = 1.27\,THz. The time transients and the corresponding spectra used to extract the transmittance are shown in Section~S1 of the Supplementary Information. Theoretically, we use a three-level $\Lambda$-type level scheme to model the EIT phenomenon. The modes of our EIT structure can be compared with that of an atomic EIT system, where $\ket{0}$ represents the ground state, $\ket{1}$ represents the excited state (which is the dipole mode of the rod resonator), and $\ket{2}$ represents the metastable state (which is the LC mode of the SRR), as shown schematically in Figure~\ref{fig1}a. The red arrows in Figure~\ref{fig1}a denote the allowed transitions, i.e., $\ket{0}-\ket{1}$ and $\ket{1}-\ket{2}$. We see that there are two possible light excitation pathways: first, the direct excitation $\ket{0}\to\ket{1}$ and second, the indirect excitation $\ket{0}\to\ket{1}\to\ket{2}\to\ket{1}$, each of which tends to produce a phase shift that would result in a net phase difference of $\pi$ between the direct and indirect excitation pathways -- the destructive interference condition that would result in the EIT-{\it like} response. It can be seen that the theoretical curve obtained using the coupled oscillator model~\cite{Haldar2025AQT} (see Section~S2 in the Supplementary Information), i.e., the red-dashed curve in Figure~\ref{fig1}b displays a very good agreement with the experimental one.

To explore the EIT nonlinearities, we perform 2D-THz spectroscopy~\cite{Raab2019OptExp,Markmann2021NaPh,Yang2023NRM,Dutta2025AFM,Wan2019PRL,Pal2021PRX,Riepl2021LSA} on the EIT metamaterial, where we use two co-propagating THz pulses, illustrated schematically in Figure~\ref{fig1}a. The two phase-locked collinear THz pulses, denoted by ${\bf E}^{\rm in}_{\rm A}(t,\tau)$ and ${\bf E}^{\rm in}_{\rm B}(t)$, are generated via optical rectification in two ZnTe crystals. Here, $t$ represent the THz sampling time (that we refer as the {\it real time}) and $\tau$ represent the delay between the arrival of the two THz pulses on the THz-EIT metamaterial. The first THz pulse induces current in the metamaterial structure and creates the EIT response, which is subsequently probed by the second THz pulse. At $\tau=0$, however, when both the THz pulses are present simultaneously, nonlinear currents are induced in the system. The nonlinear signal, ${\bf E}_{\rm NL}(t,\tau)$ is obtained by subtracting the single pulse interactions, ${\bf E}_{\rm A}(t,\tau)$ and ${\bf E}_{\rm B}(t)$, from the total response ${\bf E}_{\rm AB}(t,\tau)$, when both pulses interact with the system. Figure~\ref{fig1}c shows a representative example of the experimentally measured nonlinear response of the EIT metamaterial at a specific delay time.

\begin{figure}
    \centering
    \includegraphics[width=0.85\linewidth]{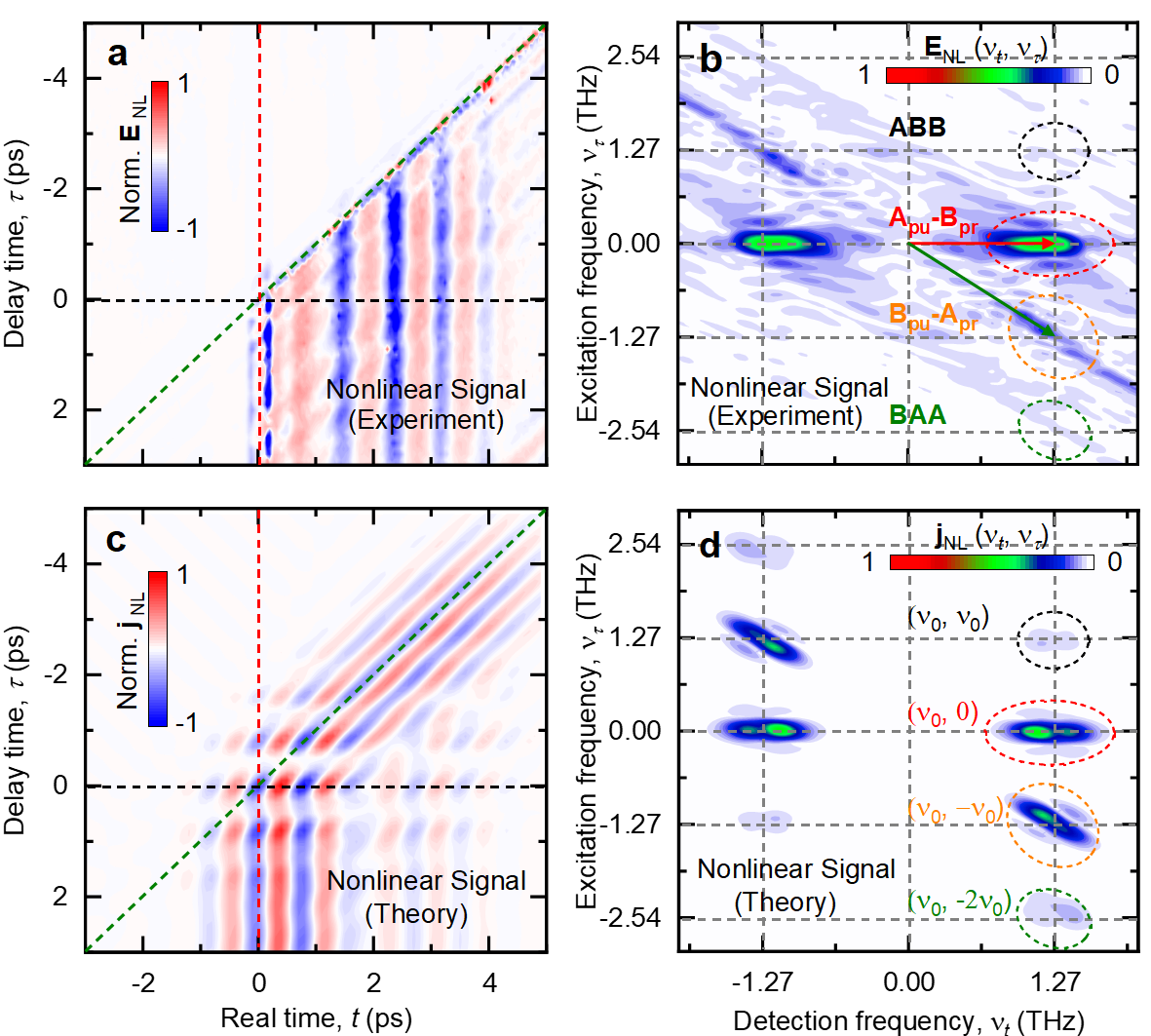}
    \caption{{\bf (a)} Contour plot of the emitted nonlinear signal ${\bf E}_{\rm NL}(t,\tau)$ from the THz EIT metamaterial. The green- and red-dashed lines correspond to the propagation wavefront of the driving THz fields A and B, respectively. The black-dashed line indicates the zero delay time. {\bf (b)} Normalized 2D Fourier spectrum of the experimental ${\bf E}_{\rm NL}(\nu_{t},\nu_{\tau})$ as a function of the detection frequency, $\nu_{t}$ and excitation frequency, $\nu_{\tau}$. The dashed ellipses indicate the positions of the nonlinear signals in the 2D frequency map. The green and red arrows are the frequency vectors associated with the THz pulses ${\bf E}_{\rm A}$ and ${\bf E}_{\rm B}$, respectively. Here, $\nu_{0} = 1.27$\,THz is the EIT peak frequency. Theoretically-modeled {\bf (c)} nonlinear current ${\bf j}_{\rm NL}(t,\tau)$ and the corresponding {\bf (d)} 2D Fourier spectrum ${\bf j}_{\rm NL}(\nu_{t},\nu_{\tau})$.}
    \label{fig2}
\end{figure}

Figure~\ref{fig2}a shows the experimentally obtained EIT nonlinear signal, ${\bf E}_{\rm NL}(t,\tau)$, plotted as functions of the real time $t$ and the delay time $\tau$. The single THz field along with the both field responses are provided in Section~S3 of the Supplementary Information. The evolution of the ${\bf E}_{\rm NL}$ in $t$ corresponds to the EIT signal at 1.27\,THz. In addition, we also see that ${\bf E}_{\rm NL}$ evolves across $\tau$ that characterizes the dependence of the nonlinear signal on the frequency components of the two incidents THz fields that evolves with $\tau$. As $\tau$ varies, the total incident field undergoes a strong modulation, resulting in significant changes in ${\bf E}_{\rm NL}$. By taking a 2D Fourier transform of ${\bf E}_{\rm NL}(t,\tau)$, we obtain the nonlinear signals in the frequency domain, ${\bf E}_{\rm NL}(\nu_t,\nu_{\tau})$, shown in Figure~\ref{fig2}b, where $\nu_t$ and $\nu_{\tau}$ are the detection and excitation frequencies, respectively. The 2D frequency spectrum contains four types of nonlinear signals (highlighted by the dashed ellipses) that are referred to as the $\chi^{(3)}$ nonlinear signals~\cite{Hamm2011CUP,Elsaesser2019IOP} because they arise from three-field interactions, involving at least one field from each THz pulses. These signals appear at the detection frequency $\nu_t$ = $\nu_0$ = 1.27\,THz, corresponding to the EIT peak frequency. All the nonlinear signals can be described as a linear combinations of the frequency vectors corresponding to the incident THz pulses, indicated by the green ($\nu_{\rm A}$) and red ($\nu_{\rm B}$) arrows~\cite{Hamm2011CUP,Elsaesser2019IOP}, see Figure~\ref{fig2}b. It is important to note that these frequency vectors directly correspond to the wavevectors $k_{\rm A}$ and $k_{\rm B}$ in the wave-vector space that is commonly used in non-collinear 2D electron spectroscopy~\cite{Hamm2011CUP}.

The two most intense nonlinear signals in the 2D frequency map are the pump-probe signals: (i) ${\rm A}_{\rm pu}- {\rm B}_{\rm pr}$ located at ($\nu_0, 0$), the nonlinear fields of which is denoted by ${\bf E}^{\rm AB}_{\rm pp}(t,\tau)$, and (ii) ${\rm B}_{\rm pu}- {\rm A}_{\rm pr}$ located at ($\nu_0, -\nu_0$), the nonlinear fields of which is denoted by ${\bf E}^{\rm BA}_{\rm pp}(t,\tau)$, see Figure~\ref{fig2}b. These signals are referred to as pump-probe signals because the sequence of THz field interactions preserves the phase information of the probe fields, while the phase evolution of the corresponding pump fields is canceled out. For example, the ${\rm A}_{\rm pu}- {\rm B}_{\rm pr}$ signal at ($\nu_0, 0$) can be written as $\nu_{\rm AB} = \nu_{\rm A} - \nu_{\rm A} + \nu_{\rm B}$, while the ${\rm B}_{\rm pu}- {\rm A}_{\rm pr}$ signal at ($\nu_0, -\nu_0$) is expressed as $\nu_{\rm BA} = \nu_{\rm B} - \nu_{\rm B} + \nu_{\rm A}$. The emission of pump-probe signals involves a sequence where the pump pulses first engage in a two-field interaction to induce a nonlinear current in the EIT metamaterial system. Subsequently, the probe pulse interacts through a single-field interaction at a delayed time to measure the emitted nonlinear current. The other two relatively weaker nonlinear signals in the 2D frequency map correspond to the echo signals: (i) the ABB photon-echo located at ($\nu_0, \nu_0$), the nonlinear fields of which is denoted by ${\bf E}^{\rm ABB}_{\rm pe}(t,\tau)$, and (ii) the BAA photon-echo located at ($\nu_0, -2\nu_0$), the nonlinear fields of which is denoted by ${\bf E}^{\rm BAA}_{\rm pe}(t,\tau)$ . In contrast to the pump-probe signals, these signals retain the phase evolution of both the interacting fields, similar to conventional photon-echo experiments~\cite{Kurnit1964PRL,Wegener1990PRA}. The ABB photon-echo signal at ($\nu_0, \nu_0$) can be expressed as $\nu_{\rm ABB} = 2\nu_{\rm B} - \nu_{\rm A}$, whereas the BAA-echo signal at ($\nu_0, -2\nu_0$) is given by $\nu_{\rm BAA} = 2\nu_{\rm A} - \nu_{\rm B}$, see Figure~\ref{fig2}b. These nonlinear signals can be scrutinized further to explore the ultrafast relaxation and the coherence timescales that dictates how long the superposition between the bright and dark modes would persist to result in the EIT transparency window under the two-pulse excitation.

To numerically model the nonlinear response of the EIT-{\it like} phenomenon in our THz metamaterial, we use a time-resolved density matrix model, where the elements of the full density matrix are evolved according to the Liouville-von-Neumann equation in the Lindblad form~\cite{Markmann2021NaPh,Jin2025JCP,Kirsanskas2018PRB}: 
\begin{equation}
   \frac{d\rho}{dt} = -\frac{i}{\hslash}[H(t),\rho] + \mathcal{L}(\rho).
   \label{EOM}
\end{equation}
Here, $\rho$ is the density matrix corresponding to the 3-level $\Lambda$-type system. To best describe our system we consider a Hamiltonian $H(t)$ that consists of the bare electronic energies and two coupling terms: (i) a structural hybridization between the excited states, and (ii) the dipole interaction with the applied THz fields. The incident THz radiation being in resonance with our metamaterial, we exploit the rotating-wave approximation, where terms in the Hamiltonian that oscillate rapidly are neglected. The system dynamics is not only described by the time evolution of $\rho$, but also the non-Hermitian contributions in the form of the Lindblad super-operator $\mathcal{L}(\rho)$ that takes care for all the decoherence effects, including population relaxation and pure dephasing. Population relaxation is included as the decay from the excited states to the ground states with rates $\gamma_1$ and $\gamma_2$. Coherences between states are damped by dephasing rates $\gamma_{\phi}$. The Lindblad operator takes the following form:
\begin{equation}
    \mathcal{L}(\rho) = \sum_{\alpha}\gamma_{\alpha}\big[\mathscr{L}_{\alpha}\rho\mathscr{L}_{\alpha}^{\dagger}-\frac{1}{2}\big\{\mathscr{L}_{\alpha}\mathscr{L}_{\alpha}^{\dagger},\rho\big\} \big]+\sum_k\gamma_{\phi}\big[\mathscr{L}_{k}\rho\mathscr{L}_k^{\dagger}-\frac{1}{2}\big\{\mathscr{L}_k\mathscr{L}_k^{\dagger},\rho\big\}\big].
    \label{Lindblad}
\end{equation}
The above expression contains two terms, corresponding to two distinct contributions from the system. The first one is population relaxation processes that are described through transition operators of the form $\mathscr{L}_{\alpha}$, which transfers population from an excited state $|k\rangle$ into a lower state $|l\rangle$ with a given decay rate. In our case, we have two transitions: $\lvert 1 \rangle \rightarrow \lvert 0 \rangle$ with a decay rate of $\gamma_1$ and  $\lvert 1 \rangle \rightarrow \lvert 2 \rangle$ with a decay rate of $\gamma_2$. The values of $\gamma_1$ and $\gamma_2$ are taken from the modeled linear transmittance spectrum shown in Figure~\ref{fig1}c. These terms ensure that as the excited-state populations diminish, the corresponding lower states are repopulated, while simultaneously suppressing any superpositions or dampening of the associated quantum coherences that involve the decaying states. In contrast, pure dephasing processes are modeled through diagonal projection operators $\mathscr{L}_k$ having a dephasing rate of $\gamma_{\phi}$, attenuating the off-diagonal terms of the density matrix without affecting populations or without exchanging the energy. Physically, these two mechanisms represent the energy relaxation occurring between the EIT quantum states and the environmental dephasing of coherences due to impurity or disorder (such as structural defects in the metamaterials) in the EIT system, respectively. Implementation of these mechanisms provide a rather realistic description of the dissipation, ensuring that the ensuing dynamics capture both coherent light-matter coupling and irreversible decoherence effects. Figure~\ref{fig2}c shows the emitted nonlinear current response while Figure~\ref{fig2}d shows the corresponding Fourier-transformed 2D frequency spectrum, which shows a very good agreement with our experiments. 

\begin{figure}
    \centering
    \includegraphics[width=\linewidth]{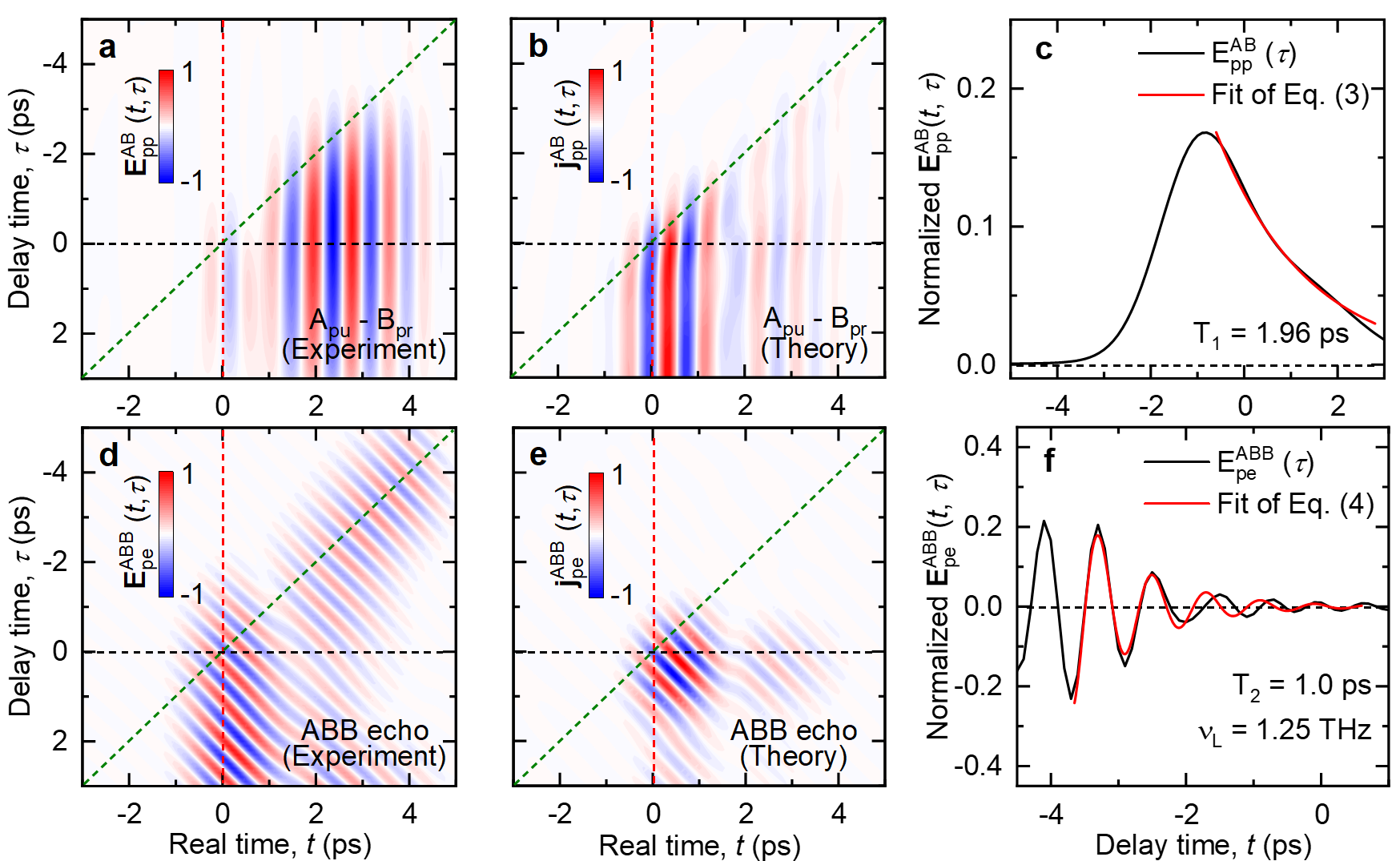}
    \caption{The 2D time-domain contour plots of {\bf(a)} experimental ${\bf E}^{\rm AB}_{\rm pp} (t,\tau)$ and {\bf(b)} theoretical ${\bf j}^{\rm AB}_{\rm pp} (t,\tau)$, obtained via inverse Fourier transform of the nonlinear ${\rm A}_{\rm pu}-{\rm B}_{\rm pr}$ signal, indicated by the red-dashed ellipses in Figures~\ref{fig2}(b,d). {\bf (c)} Temporal trace of ${\bf E}^{\rm AB}_{\rm pp}(\tau)$ at a fixed real time of $t=1.2$\,ps. The red-solid line represents the numerical fit of Equation~\ref{PPfit}. The 2D time-domain contour plots of {\bf(d)} experimental ${\bf E}^{\rm ABB}_{\rm pe} (t,\tau)$ and {\bf(e)} theoretical ${\bf j}^{\rm ABB}_{\rm pe} (t,\tau)$, obtained via inverse Fourier transform of the nonlinear ABB photon-echo signal, indicated by the black-dashed ellipse in Figures~\ref{fig2}(b,d). {\bf (f)} Temporal trace of ${\bf E}^{\rm ABB}_{\rm pe}(\tau)$ at a fixed real time of $t=4.5$\,ps. The red-solid curve represents the numerical fit of Equation~\ref{ABBfit}. The green- and red-dashed lines correspond to the propagation wavefront of the driving THz fields ${\bf E}_{\rm A}$ and ${\bf E}_{\rm B}$, respectively. The black-dashed line indicates the zero delay time.}
    \label{fig3}
\end{figure}

We selectively filter the ${\rm A}_{\rm pu}-{\rm B}_{\rm pr}$ signal in Figure~\ref{fig2}b and Fourier back-transform it in the time-domain to address the bright mode recovery dynamics of EIT structure. Figures~\ref{fig3}a and~\ref{fig3}b show the respective experimental and theoretical contour plot of the nonlinear ${\rm A}_{\rm pu}-{\rm B}_{\rm pr}$ signal in the time domain. We note that the ${\rm A}_{\rm pu}-{\rm B}_{\rm pr}$ signal is the only non-oscillatory pump-probe (PP) component of the total ${\bf E}_{\rm NL}$ along the $\tau$ direction. This signal allows us to track the decay of the current amplitude, analogous to the energy relaxation of the bright-mode after being excited by the THz pump pulse. The exponential decay of the ${\rm A}_{\rm pu}-{\rm B}_{\rm pr}$ field component along the $\tau$-axis takes the form~\cite{Riepl2021LSA}
\begin{equation}
    E_{\rm PP}(\tau) = A_{1}{\rm exp}(-\tau/T_1)
    \label{PPfit}
\end{equation}
for a fixed value of the real time (such as, $t=1.2$\,ps), with $A_1$ denoting the signal amplitude and $T_1$ being the bright-mode relaxation time. As shown in Figure~\ref{fig3}c, we find that the bright-mode energy relaxation (analogous to the current amplitude decay) $T_1 = 1.96 \pm 0.11$\,ps. This represents the characteristic timescale over which the radiative energy, stored in the bright dipole, dissipates through ohmic loss and radiative re-emission.

The nonlinear ABB photon-echo signal, in contrast, is characterized by oscillations along the delay axis ($\tau$), arising from the coherent interference between the bright and the dark resonator currents. This oscillatory ABB photon-echo signal exhibits a sinusoidal behavior along the delay ($\tau$) axis, described by~\cite{Riepl2021LSA}, 
\begin{equation}
    E^{\rm ABB}_{\rm pe}(\tau) = A_{2}{\rm sin}(2\pi\nu_{\rm L}\tau + \phi){\rm exp}(-\tau/T_2)
    \label{ABBfit}
\end{equation}
where, $A_2$ and $\phi$, respectively, represent the amplitude and phase of the oscillations with a frequency of $\nu_{\rm L}$, corresponding to the frequency of ABB photon-echo. Fitting this expression to the experimental data at a fixed value of the real time (such as, $t=4.5$\,ps, we obtain $T_2=1.0\pm 0.1$\,ps, which characterizes the lifetime of the coherent energy exchange between the bright and dark modes resonators. On analyzing the BAA photon-echo signal, we find identical values of $T_2$, see Section~S5 of the Supplementary Information. The coherence allows the destructive interference between the radiative and subradiant modes, leading to the formation of the EIT transparency window in our metamaterial system. We note that the coherence time (i.e., $T_2$) is smaller than the population relaxation time ($T_1$). This inequality, in fact, reflects that the phase coherence of our system is typically lost before the system has completely returned to its equilibrium state, i.e., before the energy or amplitude of the current has fully decayed. The inequality can be described using $1/T_2 = 1/(2T_1) + 1/T_{\phi}$~\cite{Deng2017PRB}, where $T_{\phi}$ denotes the pure dephasing time that accounts for processes such as inhomogeneous broadening, radiative damping, and phase randomization. From our experiments, we obtain $T_{\phi}$ to be 1.34\,ps, consistent with the $\gamma_{\phi}$ ($= 0.8$\,ps$^{-1}$) used in our model. Despite the energy due to the dipole remaining in the bright resonator for several picoseconds, the phase correlation between the bright and dark resonators decays faster due to the dominant influence of these dephasing mechanisms. This interplay between recovery and coherence loss ultimately dictates the shape and width of the EIT-{\it like} window.

\begin{figure}
    \centering
    \includegraphics[width=0.9\linewidth]{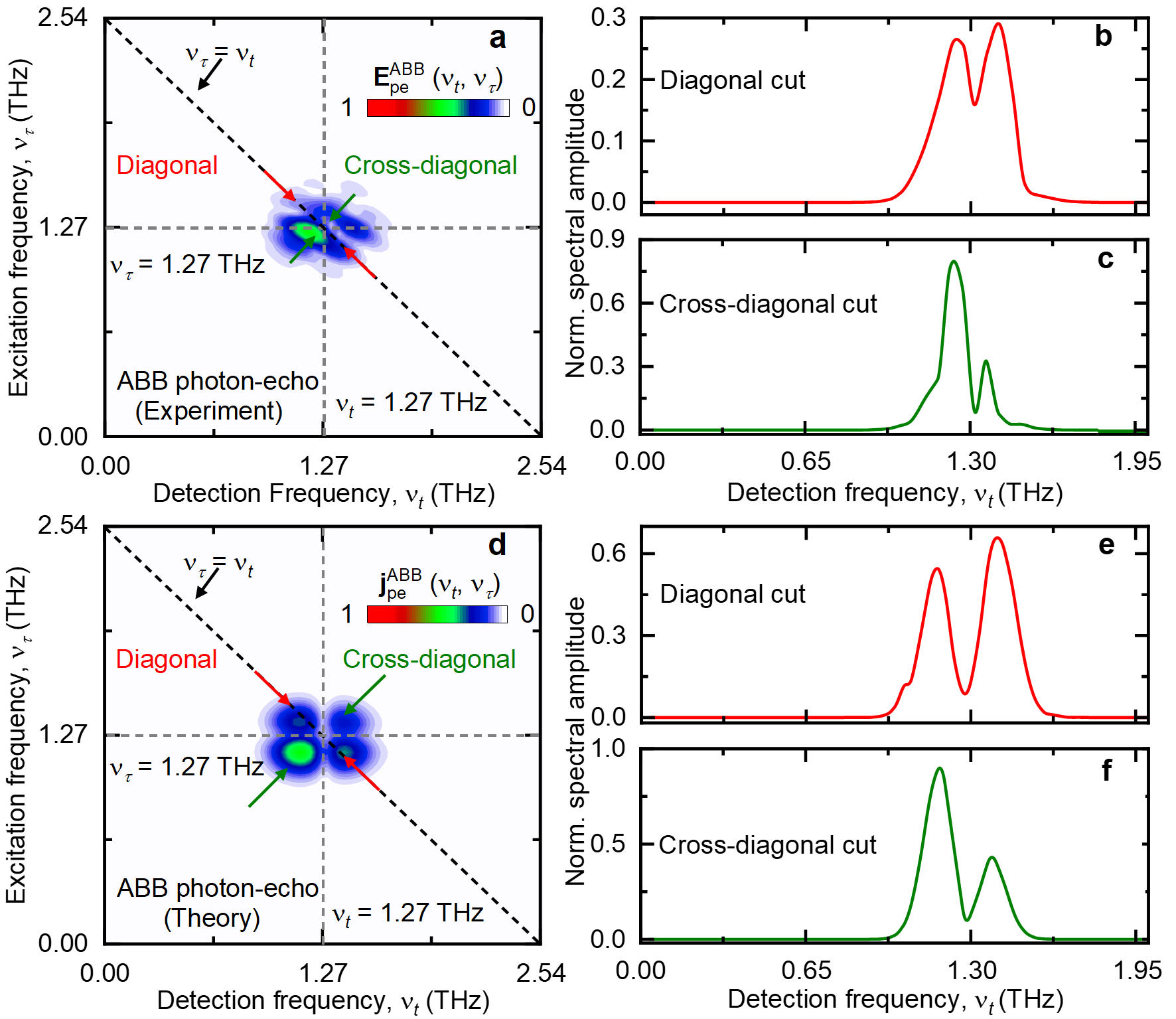}
    \caption{{\bf(a)} The experimental nonlinear ABB photon-echo signal showing a multi-peak nature that results from the EIT-{\it like} phenomenon. {\bf(b)} The diagonal and {\bf(c)} the cross-diagonal spectral slices as indicated by the arrows in (a). {\bf(d)} The theoretical nonlinear ABB photon-echo signal showing a multi-peak nature that remarkably reproduces the experiments. {\bf(e)} The diagonal and {\bf(f)} the cross-diagonal spectral slices as indicated by the arrows in (d). The black-dashed lines in (a) and (d) along the diagonal indicates the $\nu_{\tau}=-\nu_{t}$ line. The horizontal and vertical gray-dashed lines in (a) and (d) represent the $\nu_{\tau}=1.27$\,THz and $\nu_{t}=1.27$\,THz lines, respectively.}
    \label{fig4}
\end{figure}

In a $\Lambda$-type system, similar to the one considered in our work, the control field creates two dressed states (namely the upper and the lower dressed states). Interference between the dressed states with the probe or the control transition provides the necessary pathways for the destructive interference, eventually leading to the observation of the transparency window. Taking a closer look at the spectral shape of the ABB photon echo in the 2D frequency domain, we note that the signal has a multi-peak nature. Precisely, the ABB signal has four peaks -- a {\it hallmark} signature of the formation of dressed states~\cite{Deng2024PRA,Liu2020JPCL,Jin2025JCP}. The multi-peak nature observed in our experimental ABB photon echo signal in Figure~\ref{fig4}a is remarkably reproduced by our theoretical model as shown in Figure~\ref{fig4}d. The diagonal and the cross diagonal slices obtained from experiments (Figures~\ref{fig4}b and~\ref{fig4}c) match well the ones obtained from the theoretical model (Figures~\ref{fig4}e and~\ref{fig4}f), revealing the lower dressed state located at $\nu_t<1.27$\,THz and the upper dressed state located at $\nu_t>1.27$\,THz, with a spectral hole at $f_{\rm EIT}=1.27$\,THz. We note that the spectral linewidths along the cross-diagonal is smaller than that along the diagonal. The spectral width along the diagonal line indicates the inhomogeneous broadening that may result from the inhomogeneity in the metamaterial structures itself and is a cumulative effect, while the width along the cross-diagonal line represents the homogeneous broadening, resulting from the natural lifetime broadening which is otherwise affects the system in a similar manner. 

\section{Conclusion}
In conclusion, we have explored the nonlinearities in a THz-EIT metamaterial using 2D-THz spectroscopy, substantiated  by a time-resolved density matrix model according to the Liouville-von-Neumann equation in the Lindblad form. From the temporal decay of the nonlinear pump-probe signal we extract the relaxation time of the excited state in our EIT-{\it like} system, while the temporal variation of the four-wave mixing signals allowed us to unveil the timescales for coherent transfer of energy between the bright and the dark resonators. We found that the relaxation timescale is almost twice that of the coherence timescales. The coexistence of picosecond dephasing with a longer population relaxation time confirms that the metamaterial operates in a moderately coherent regime where strong near-field coupling sustains the transient quantum interference, despite the presence of dissipation. These findings provide a comprehensive picture of the interplay between the current amplitude decay and the coherence dynamics in artificial $\Lambda$-type THz-EIT systems, revealing the distinct nonlinear temporal signatures of bright-dark mode coupling that underpins the coherent control of the transparency in rather classical metamaterial platforms.

\clearpage

\clearpage
\section{Methods Section}
{\it Nonlinear 2D THz experiments:}
The nonlinear 2D THz experiments are carried out by generating two co-propagating phase-locked THz pulses, ${\bf E}^{\rm in}_{\rm A}$ and ${\bf E}^{\rm in}_{\rm B}$, separated by the delay time, $\tau$. Both pulses are generated via optical rectification of two phase-locked 100-fs pulses at a central wavelength of 800\,nm in two 0.5\,mm-thick ZnTe crystals oriented along (110). The arrival of the two THz pulses on the sample is precisely controlled by adjusting a delay stage. The two pulses interact with the sample in a collinear geometry, ensuring that all nonlinear signal components appear simultaneously within the 2D frequency map. Both pulses are measured using free-space electro-optic sampling, which is carried out in a 0.5\,mm-thick (110)-oriented ZnTe crystal, which is optically bonded on a 2\,mm-thick (100)-oriented ZnTe crystal. The nonlinear THz field ${\bf E}_{\rm NL}(t,\tau)$ is determined using the relation ${\bf E}_{\rm NL}(t,\tau) = {\bf E}_{\rm AB}(t,\tau) - {\bf E}_{\rm A}(t,\tau) - {\bf E}_{\rm B}(t)$, where ${\bf E}_{\rm AB}(t,\tau)$ is the transmitted field when both pulses interact with the sample, while ${\bf E}_{\rm A}$ and ${\bf E}_{\rm B}$ are the transmitted fields obtained when individual pulses interact with the sample. The individual pulse strengths are kept such that $|{\bf E}_{\rm B}| > |{\bf E}_{\rm A}|$. All experiments are carried out at room temperature in an inert N$_{2}$ atmosphere.

\noindent{\it Theoretical model:}
We consider a three-level $\Lambda$-type system~\cite{Liu2017Naph}, where the ground state is represented by $\ket{0}$, the bright excited state is represented by $\ket{1}$ and the dark excited state is $\ket{2}$. While the $\ket{0}\leftrightarrow\ket{1}$ excitation is dipole-allowed and leads to a strong absorption, the $\ket{0}\leftrightarrow\ket{2}$ is a dipole-forbidden transition. The incident THz pulse directly couples to the $\ket{0}\leftrightarrow\ket{1}$ transition. The $\ket{1}\leftrightarrow\ket{2}$ transition, in contrast, results from the introduction of the coupling field (also called the control field) through a structural hybridization parameter $g_{\rm s}$. Effectively, we have a dipole coupling between the ground and the bright state and a hybridization (coupling) between the bright and the dark states. The Hamiltonian that best represents our case would consist of the bare electronic energies $(H_0)$ and the coupling terms: A structural hybridization between the excited states $(H_{\rm c})$, and the dipole interaction with the applied THz fields $(H_{\rm int})$ and can be expressed as $H(t) = H_0 + H_{\rm c} + H_{\rm int}(t)$~\cite{Huang2022RP}, where the bare electronic Hamiltonian is given by $H_0 = E_0 |0\rangle \langle0| + E_1 |1\rangle \langle1| + E_2 |2\rangle \langle2|$, the structural hybridization between the excited states is given by $ H_{\rm c} = g_{\rm s} \left( |1\rangle \langle 2| + |2\rangle \langle 1| \right)$, and the dipole interaction with the external electromagnetic field $\mathbf{E}(t)$ is expressed as $H_{\text{int}}(t) = - \mathbf{d} \cdot \mathbf{E}(t) = - \mu_{01} E(t) \left( |0\rangle \langle 1| + |1\rangle \langle 0| \right)$. The expression for ${\bf E}(t)$ used in our model that best reproduces our experiments is given by~\cite{Dutta2025AFM}
\begin{equation}
    \mathbf{E(t)} \;=\; E_0 \, 
    \cos \!\left[ \, 2\pi f_0\,\gamma \left( e^{\tfrac{t-t_0}{\gamma}} - 1 \right) \,\right] \,
    \exp \!\left( -\frac{(t-t_0)^2}{2\sigma^2} \right)\mathbf{\hat{x}},
\end{equation}
where $E_0$ is the field amplitude, $t_0$ represents the temporal center of the pulse, and $f_0$ is the central (carrier) frequency of the pulse. $\gamma$ is the chirp parameter that controls the exponential variation of frequency with time. $\sigma$ is the Gaussian pulse width that sets the temporal envelope. All parameters corresponding to the THz pulse are provided in the Supplementary Information, see Section~S6. 

In our model, we set the initial condition where the system is considered to be in the ground state, i.e., $\rho(t=0) = \ket{0}\bra{0}$. To compute $\rho(t)$, the equations of motion (as per Eq.~\ref{EOM}) are integrated numerically using a fourth-order Runge-Kutta process. This is done to ensure a resolution that captures most of the experimental features and at the same time economical for computational power available with us. At each timestep, the instantaneous Hamiltonian is constructed from the bare energies, the structural coupling and the external field. The commutator $[H,\rho]$ is then evaluated and subsequently the density matrix is updated. The measurable output in our case is the current $j(t)$, which is expressed as:
\begin{equation}
    j(t) = \frac{Ne}{m}Tr[\rho(t)\hat{p}]
    \label{current}
\end{equation}
where, $N$, $e$ and $m$ are the electron density, charge and mass, respectively. Note that $\hat{p}$ is the momentum operator, which is related to the dipole operator with its time-derivative, i.e.,
\begin{equation}
    \hat{p} = \frac{m}{e}\frac{d}{dt}\hat{d}.
    \label{momentum}
\end{equation}
Considering that the dipole operator $\hat{d}$ has no explicit time dependence, for the unperturbed Hamiltonian $H_0$, the Heisenberg equation gives
\begin{equation}
    \frac{d}{dt}\hat{d} 
    = \frac{i}{\hslash} \big[ H_0 , \hat{d} \big] = \frac{i}{\hslash} (E_k - E_l) d_{kl}.
    \label{heisenberg}
\end{equation}
Using Eqs.~(\ref{current}-\ref{heisenberg}), we have,
\begin{equation}
    j(t) = Tr\big[\rho(t)\frac{d}{dt}\hat{d}\big]=i\frac{N}{\hslash}\sum_{kl}\rho_{kl}(t)(E_k - E_l) d_{kl}.
    \label{Fcurrent}
\end{equation}
Using Eq.~\ref{Fcurrent}, we have evaluated the currents corresponding to the single THz pulse interactions, i.e., $j_{\rm A}(t,\tau)$ and $j_{\rm B}(t)$ and also for both pulses simultaneously present in our EIT-like metamaterial, i.e., $j_{\rm AB}(t,\tau)$. Note that in our calculations, we have considered both $N$ and $\hslash$ to be 1, i.e., normalized units. The nonlinear is then evaluated using $j_{\rm NL}(t,\tau) = j_{\rm AB}(t,\tau) - j_{\rm A}(t,\tau) - j_{\rm B}(t)$. The nonlinear current response (shown in Figure~\ref{fig2}c) is then Fourier-transformed with respect to both the real time ($t$) and the delay time $(\tau)$, yielding the 2D frequency-domain response $j_{\rm NL}(\nu_t, \nu_{\tau})$, shown in Figure~\ref{fig2}d. The resultant nonlinear 2D frequency spectrum contains all the distinct signals separated in the frequency space, as obtained in our experiments.

\begin{addendum}
\item[Author Contributions]
All authors contributed to the discussion and interpretation of the experiment and to the completion of the manuscript. A.H. designed and fabricated the metamaterials and performed the experiments. A.H., under the supervision of S.S.P. characterized the metamaterials. A.H. analyzed the data and developed the theoretical model. S.P. conceived and supervised the project. A.H. and S.P. drafted the manuscript.

\item[Acknowledgements]
A.H. and S.P. acknowledge the support from DAE through the project Basic Research in Physical and Multi-disciplinary Sciences via RIN4001 and and New Frontiers in Earth, Atmospheric, Planetary, Stellar Physics, Material Sciences and Rare Event Searches via RNI4011. S.P. additionally acknowledges the start-up support from DAE through NISER. S.S.P. acknowledges the funding support from DAE through the project RTI4003. The authors acknowledge Satyaprasad P. Senanayak for availing the metal evaporation facility and Kush Saha for the fruitful discussions on the theoretical model.

\item[Competing Interests] The authors declare that they have no competing financial interests.

\item[Data Availability] {The datasets used in the current study available from corresponding authors upon reasonable request.}

\item[Code Availability] {The codes associated with this study are available from the corresponding authors upon reasonable request.}

\item[Correspondence]{All correspondence should be addressed to A.H. (amit.haldar@niser.ac.in) or S.P. (email: shovon.pal@niser.ac.in)}
\end{addendum}

\end{document}


\maketitle
\begin{affiliations}
\item School of Physical Sciences, National Institute of Science Education and Research, An OCC of HBNI, Jatni, 752 050 Odisha, India
\item Department of Condensed Matter Physics and Materials Science, Tata Institute of Fundamental Research, 400 005 Mumbai, India
\end{affiliations}

\begin{abstract}
This supplementary information contains details on the metamaterial structure, the transmitted THz electric field from the substrate, the control-field dependence in the metamaterial, and the 2D pump-probe and echo signals. and the 2D pump-probe and echo signals. The contents are sectionized as:\\
\textbf{S1. Metamaterial and THz field.}\\
\textbf{S2. Coupled Oscillator model and control field dependence.}\\
\textbf{S3. Single and dual THz field responses.}\\
\textbf{S4. Pump-probe signal}\\
\textbf{S5. BAA photon-echo signal}\\
\textbf{S6. Modeled THz input pulse}\\
\end{abstract}

\maketitle

\section{Metamaterial and THz spectrum}
\begin{figure}[bh!]
    \centering
    \includegraphics[width=0.98\linewidth]{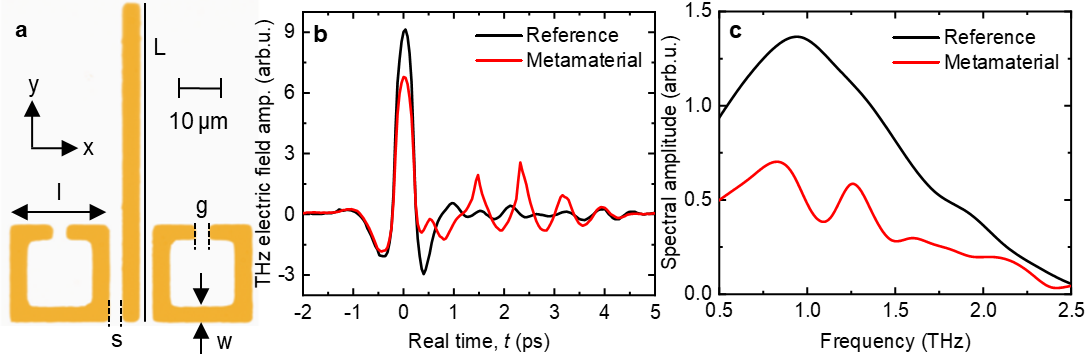}
    \caption{{\bf(a)} Optical microscopic image showing the unit cell of EIT meta-structure. The EIT-{\it like} metamaterial unit cell consists of a central rod resonator accompanied by a pair of split-ring resonators (SRRs) positioned symmetrically on either side of the rod. The geometric parameters of the metamaterial structure are as follows: $L = 75$\,$\mu$m, $l = 23.5$\,$\mu$m, $w = 5$\,$\mu$m, $g = 3$\,$\mu$m, $s = 3$\,$\mu$m with periodicity of 67\,$\mu$m and 93\,$\mu$m, along the $x$- and $y$-directions, respectively. {\bf(b)} THz electric-field transients transmitted through the substrate and the metamaterial sample and {\bf(c)} the corresponding THz spectra.}
    \label{Sfig1}
\end{figure}

We fabricated the metamaterial structures on a THz-transparent, $z$-cut quartz substrate of thickness 0.5\,mm, using maskless lithography. The fabrication procedure is detailed elsewhere~\cite{Haldar2025AQT}. Figure~\ref{Sfig1}a shows the optical microscopic image of the fabricated unit cell of the EIT meta-structure. Figure~\ref{Sfig1}b show the transmitted time transients through the substrate (bare $z$-cut quartz) and metamaterial sample. Note that the incident THz electric field is polarized along the $y$ direction, i.e., along the rod. The corresponding THz spectra obtained by taking the Fourier transforms of the time transients are shown in Figure~\ref{Sfig1}c.

\section{Coupled Oscillator model and control field dependence}
We use two-coupled, three-level system to model our hybrid phonon EIT-like system. We start with the EIT-like structure, where the rod interacts directly with the input electromagnetic field, allowing an electric-dipole transition, and serves as the bright mode. In addition, the SRR pair subsequently gets excited through the $x$-component of the THz electric field induced by the rod and acts as the dark mode. In this scenario, the field induced by the dipole excitation of the rod destructively interferes with the capacitance gap of the SRR, thereby opening up a transparency window. While  the extended coupled oscillator (ECO) model provides a more general description of radiative interference~\cite{Lovera2013ACSNano}, we adopt the conventional coupled oscillator (CCO) model for our EIT-like system. This adaptation is appropriate because, in our EIT-like structure, only the bright mode (i.e., the rod) couples significantly to the external electromagnetic field. In contrast, the dark mode (i.e., the SRRs) is effectively shielded from direct excitation. The simplified CCO model accurately captures the key features of the destructive interference and the transparency without requiring additional complexity. Within the framework of CCO, the destructive interference can be modeled using the following equations of motion for the bright and dark oscillators:
\begin{align}
    \ddot{x}_b(t) + \gamma_b\dot{x}_b(t) + \omega_b^2x_b(t) + \kappa^2x_d(t)  &= gE_0, \\
    \ddot{x}_d(t) + \gamma_d\dot{x}_d(t) + \omega_d^2x_d(t) + \kappa^2x_b(t) &= 0.
\end{align}
Here, $x_{\rm b}$ and $x_{\rm d}$ are the amplitudes of the bright (rod) and dark (SRR) mode resonators. $\omega_{\rm b}$ and $\omega_{\rm d}$ are the resonance frequencies and $\gamma_{\rm b}$ and $\gamma_{\rm d}$ are the damping rates of the bright and dark resonators, respectively. $g$ is the coupling coefficient that mediates the coupling of the bright mode with the incident electromagnetic field and $\kappa$ is the coupling coefficient between the bright mode and dark mode resonators. The transmission for the proposed EIT-like system can be expressed using the equation~\cite{Ling2018Nanoscale}:
\begin{align}
    {\rm T} = 1 - \left|\frac{g(\omega - \omega_{\rm d} + \textit{i}\gamma_{\rm d})}{(\omega - \omega_{\rm b} + \textit{i}\gamma_{\rm b})(\omega - \omega_{\rm d} + \textit{i}\gamma_{\rm d})+\kappa^2}\right|^2.
\end{align}
\begin{figure}[b!]
    \centering
    \includegraphics[width=0.9\linewidth]{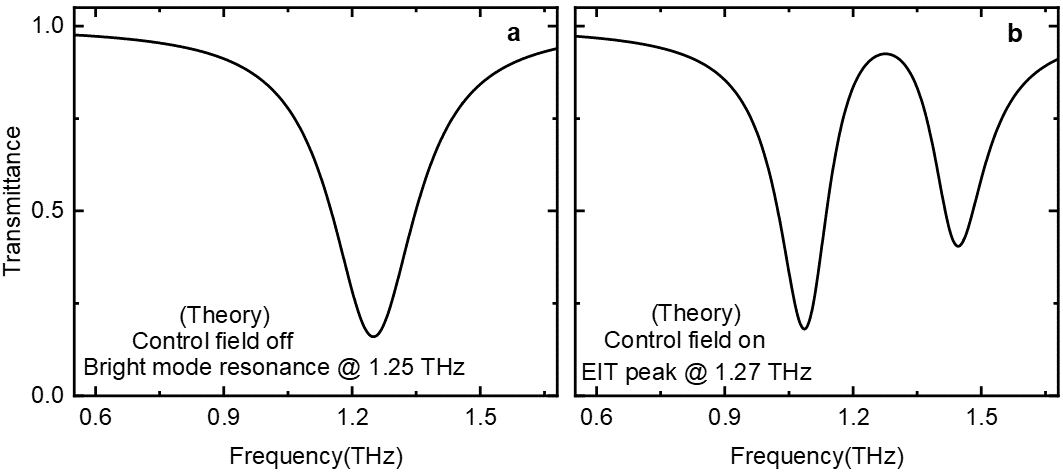}
    \caption{Theoretically calculated transmittance spectra of EIT-{\it like} THz metamaterial {\bf(a)} in the absence of control/coupling field and {\bf(b)} in the presence of control/coupling field. In presence of the control/coupling field, we observe the EIT peak at 1.27\,THz, which reproduces our experimental observation in Figure~1b of the main manuscript.}
    \label{Sfig2}
\end{figure}
Figures~\ref{Sfig2}a and~\ref{Sfig2}b show the theoretical plots of the EIT metamaterial structure in the absence and presence of the control field, respectively. When the control field is turned off, the transmittance spectrum shows a resonance dip at 1.25\,THz. This corresponds to the dipole resonance of the rod, being directly excited by the incident THz pulse polarized along its length and hence behaves as the bright mode. In contrast, when the control field is applied, the $x$-component of the THz electric field generated by the rod near its resonance excites the SRR pair. The field associated with the rod's dipole excitation interferes destructively with the field across the SRR capacitance gap, thereby suppressing the rod's response. This suppression opens a transparency window, producing an EIT-{\it like} feature at 1.27\,THz, as shown in Figure~\ref{Sfig2}b.
\begin{figure}[b!]
    \centering
    \includegraphics[width=1.0\linewidth]{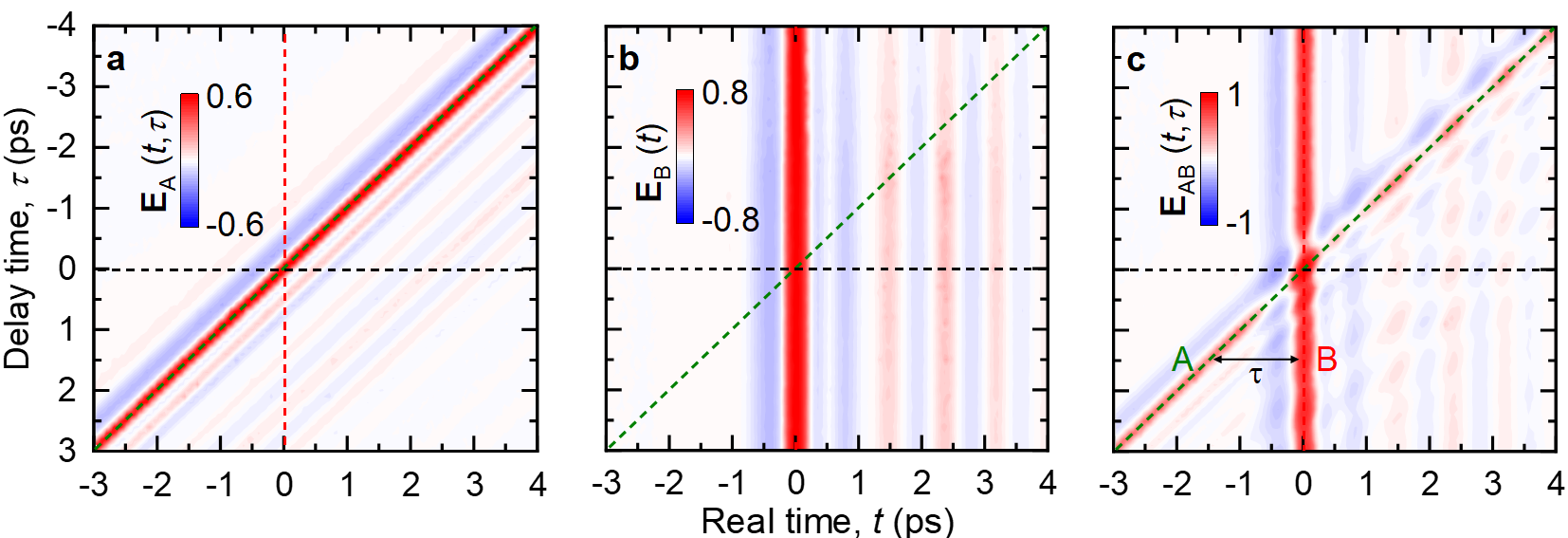}
    \caption{2D contour plots of {\bf(a)} ${\bf E}_{\rm A}(t,\tau)$, and {\bf(b)} ${\bf E}_{\rm B}(t)$ transmitted individually through the sample. {\bf(c)} Contour plot of ${\bf E}_{\rm AB}(t,\tau)$ when both THz pulses are simultaneously transmitted through the sample. The THz electric fields are normalized to the electric field value of the ${\bf E}_{\rm AB}(0,0)$. The green- and red-dashed lines correspond to the propagation wavefront of the driving THz fields $E_{\rm A}$ and $E_{\rm B}$, respectively. The black-dashed line indicates the zero delay time.}
    \label{Sfig3}
\end{figure}

\section{Single and dual THz field responses}
The experimental setup of the 2D THz spectroscopy is discussed in the methods section of the main manuscript. All the experiments are carried out at room temperature and in an inert nitrogen-purged atmosphere. Figures~\ref{Sfig3}a and~\ref{Sfig3}b show the 2D contour plots of the emitted individual responses, i.e., ${\bf E}_{\rm A}(t,\tau)$ and ${\bf E}_{\rm B}(t)$ corresponding to the THz pulses A and B, respectively. The two input THz pulses, namely ${\bf E}^{\rm in}_{\rm A}(t,\tau)$ and ${\bf E}^{\rm in}_{\rm B}(t)$ interact with the EIT metamaterial, resulting in the observed FID signals. The 2D contour plot of the total response of the sample, i.e., ${\bf E}_{\rm AB}(t,\tau)$ in the presence of both THz fields is shown in Figure~\ref{Sfig3}c, while the emitted nonlinear signal is shown in Figure~2a of the main manuscript. The nonlinear signal is obtained by subtracting the individual field responses from both field responses.

\begin{figure}[b!]
    \centering
    \includegraphics[width=\linewidth]{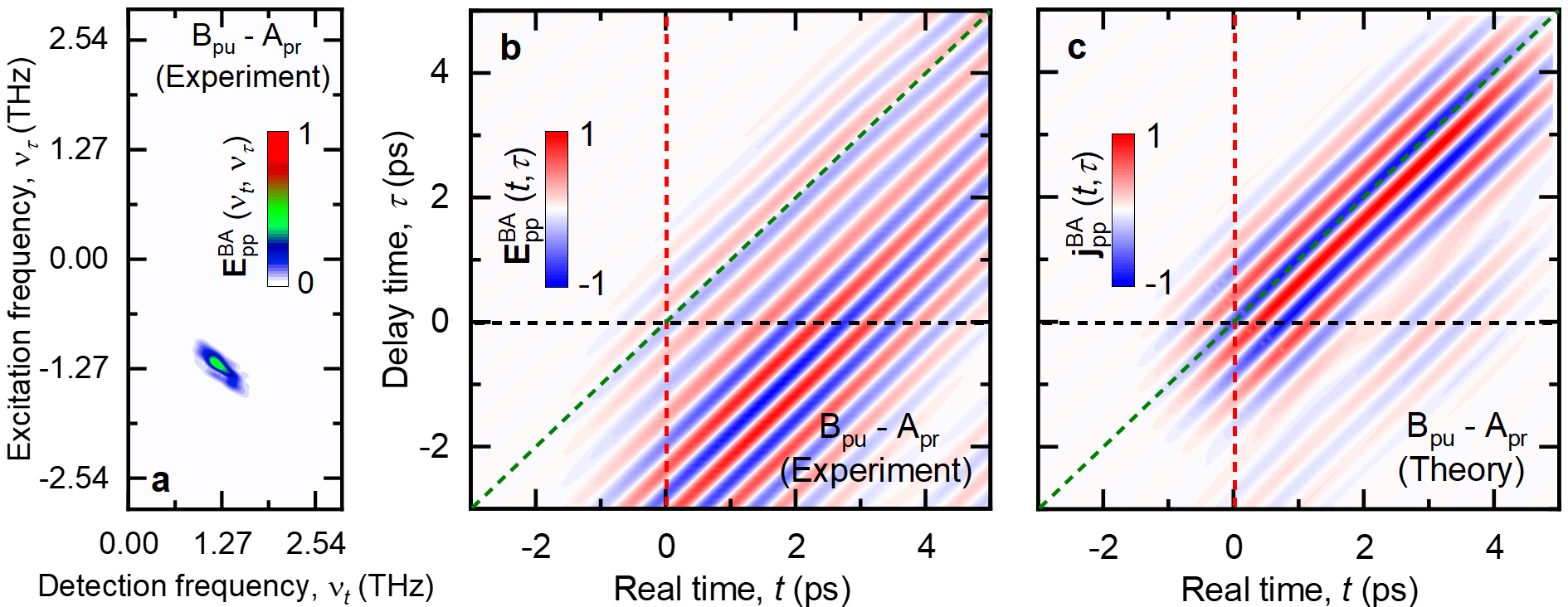}
    \caption{The 2D filtered {\bf (a)} B$_{\rm pu}$ - A$_{\rm pr}$ signals. {\bf (b)} Experimental ${\bf E}^{\rm BA}_{\rm pp} (t,\tau)$ and {\bf (c)} theoretical ${\bf j}^{\rm BA}_{\rm pp}(t,\tau)$ 2D contour plots of the inverse Fourier transformed signals, respectively. Here, the green and red-dashed lines correspond to the propagation wavefront of the driving THz fields ${\bf E}_{\rm A}$ and ${\bf E}_{\rm B}$, respectively. Meanwhile, the black-dashed line indicates the zero delay time.}
    \label{Sfig4}
\end{figure}

\section{Pump-probe signals}
In the nonlinear 2D frequency map, two strong pump-probe signals (${\rm A}_{\rm pu}-{\rm B}_{\rm pr}$ and ${\rm B}_{\rm pu}-{\rm A}_{\rm pr}$) are clearly observed. These nonlinear responses being well separated, we apply a 2D Gaussian filter to isolate and extract each component individually without the loss of information. Figure~\ref{Sfig4}a shows the ${\rm B}_{\rm pu}-{\rm A}_{\rm pr}$ signal, while the ${\rm A}_{\rm pu}-{\rm B}_{\rm pr}$ nonlinear signal is shown and discussed in the main manuscript. We perform an inverse Fourier transform to selectively reconstruct the 2D temporal contour map. While Figure~\ref{Sfig4}b shows the experimental contour plot of these signals Figure~\ref{Sfig4}c shows the corresponding theoretical 2D contour plot. Both the experimental and theoretical signals display a very good agreement with each other.

\begin{figure}[b!]
    \centering
    \includegraphics[width=\linewidth]{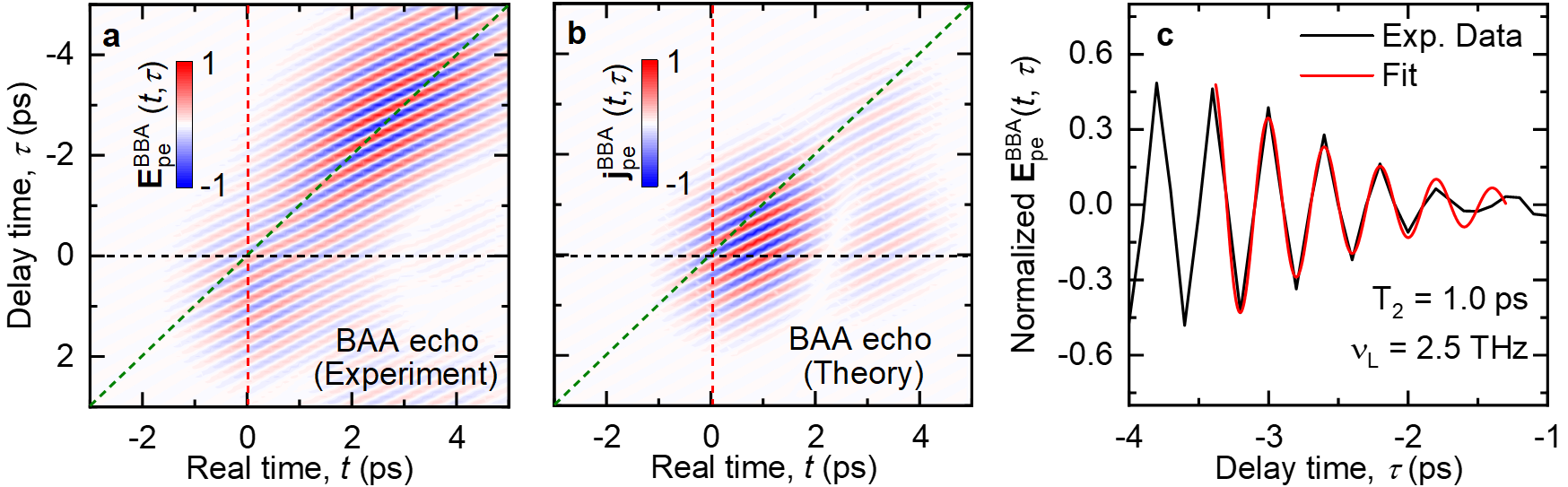}
    \caption{{\bf (a)} Experimental ${\bf E}^{\rm BAA}_{\rm pe} (t,\tau)$ and {\bf (b)} theoretical ${\bf j}^{\rm BAA}_{\rm pe}(t,\tau)$ 2D contour plots of the inverse Fourier transformed signals, respectively. Here, the green- and red-dashed lines correspond to the propagation wavefront of the driving THz fields ${\bf E}_{\rm A}$ and ${\bf E}_{\rm B}$, respectively. The black-dashed line indicates the zero delay time. {\bf (c)} The temporal trace of ${\bf E}^{\rm BAA}_{\rm pe} (t,\tau)$ at a fixed real time of $t=4.5$\,ps. The red-solid curve depicts the numerical fit of the Equation.~(4) of the main manuscript.}
    \label{Sfig5}
\end{figure}

\section{BAA photon-echo signals}
Figure~\ref{Sfig5}a shows the inverse Fourier transform of the selectively-filtered nonlinear BAA photon-echo signal as a function of $t$ and $\tau$. The nonlinear ABB photon-echo signal is shown and discussed in the main manuscript. Figure~\ref{Sfig5}b shows the theoretical 2D contour plot of nonlinear BAA photon-echo signal. Figure~\ref{Sfig5}c shows a temporal trace of ${\bf E}^{\rm BAA}_{\rm pe}(\tau)$ at a fixed value of the real time, such as $t$ = 4.5\,ps, and fitted by Equation~(4) of the main manuscript. From the fitting, we obtain $T_2=1.0\pm0.1$\,ps, which is identical with the $T_2$ of ABB photon-echo signal, discussed in the main manuscript. 
\begin{figure}[b!]
    \centering
    \includegraphics[width=0.9\linewidth]{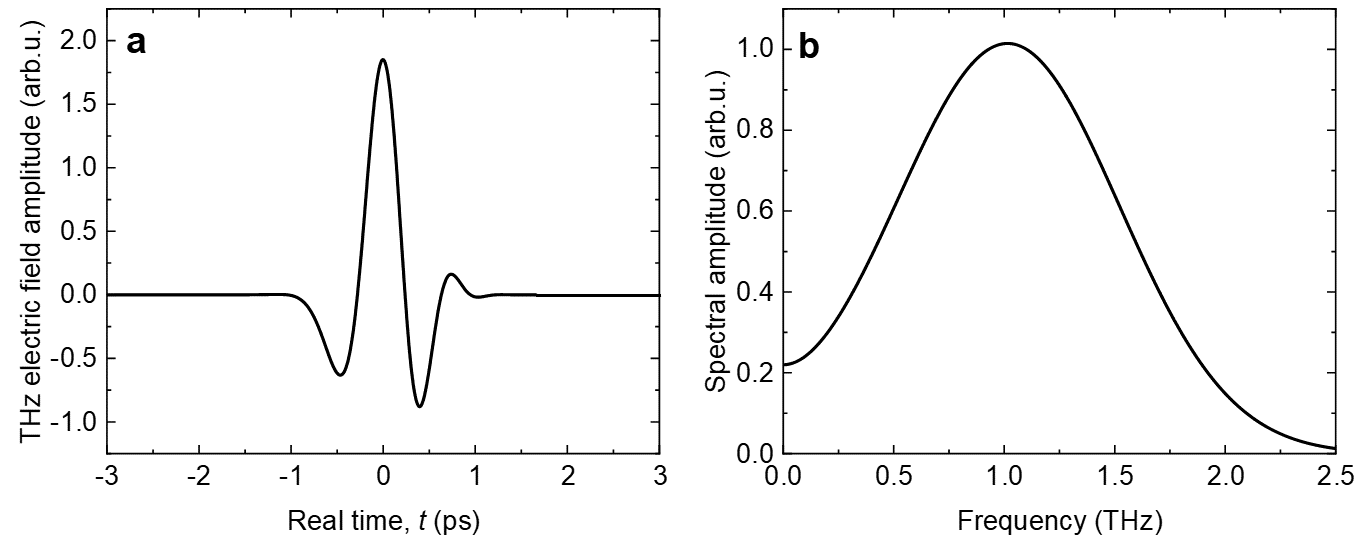}
    \caption{{\bf(a)} Modeled time transients of the input THz pulse and {\bf(b)} the corresponding spectrum of the input pulse.}
    \label{Sfig6}
\end{figure}

\section{Modeled THz input pulse}
We used the following expression to model the input THz pulses: 
\begin{equation}
E(t) = E_0 \, \cos\!\left[ 2\pi f_0 \gamma \left(\exp\!\left(\frac{t - t_0}{\gamma}\right) - 1 \right) \right] \exp\!\left[-\frac{(t - t_0)^2}{2\sigma^2} \right].
\end{equation}
where, $f_0$ is taken to be the central frequency of the THz pulse at 1.0\,THz. $\sigma=0.35$\,ps represents the Gaussian temporal width of the pulse envelope and $\gamma=1.6$\,ps is the chirp time of the pulse. $E_0$ is the amplitude of the THz electric field and $t_0$ is the temporal offset of the pulse center. We have designed our model THz pulse that closely matches our experiment, see Figure~\ref{Sfig6}.